\documentclass[a4paper]{spie}  
\usepackage[]{graphicx}

\title{The Near-Infrared Gatherer of Helium Transits (NIGHT)}

\author{Casper Farret Jentink\supit{a}, Francesco Pepe\supit{a}, Christophe Lovis\supit{a}, Sébastien Bovay\supit{a}, François Wildi\supit{a}, Bruno Chazelas\supit{a}, Michaël Sordet\supit{a},  Étienne Artigau\supit{b}, René Doyon\supit{b}, Frédérique Baron\supit{b}, Vincent Bourrier\supit{a}, Romain Allart\supit{b} and François Cochard
\skiplinehalf
\supit{a}Observatoire Astronomique de l’Université de Genève, Chemin Pegasi 51b, 1290 Versoix, Switzerland; \\
\supit{b}Département de Physique, Observatoire du Mont-Mégantic and Institut Trottier de Recherche sur les Exoplanètes, Université de Montréal, Montréal, Québec, H3T 1J4, Canada}

\authorinfo{Further author information: (Send correspondence to C.F.J.)\\C.F.J.: E-mail: casper.farret@unige.ch, Telephone: +41 (0)2 23 79 22 92\\F.P.: E-mail: francesco.pepe@unige.ch, Telephone: +41 (0)2 23 79 23 96}

\begin{document} 
  \maketitle 

\begin{abstract}
This paper provides a comprehensive overview of the subsystems of the NIGHT instrument. NIGHT (the \textbf{N}ear \textbf{I}nfrared \textbf{G}atherer of \textbf{H}elium \textbf{T}ransits) is a narrowband, high-resolution spectrograph, marking the first dedicated survey instrument for exoplanetary atmosphere observations. Developed through a collaboration between the Observatory of Geneva, several other Swiss institutes, and the Universit\'e de Montr\'eal, NIGHT aims to conduct an extensive statistical survey of helium atmospheres around 100+ exoplanets over several years. The instrument will report new detections of helium in exoplanet atmospheres and perform temporal monitoring of a subset of these.

NIGHT measures absorption from the metastable helium state during exoplanet transits, observable in a triplet of lines around 1083.3\,nm. The instrument comprises a vacuum enclosure housing the spectrograph, a front end unit for fiber injection at the telescope's focal plane, and a calibration and control rack containing calibration light sources and control hardware.

The spectrograph is optimized for efficiency around the helium triplet, achieving a throughput of approximately 71\%, uniform across wavelength and polarization. The primary disperser employs a volume-phase-holographic grating in a unique double-pass configuration, enabling a spectral resolution of 75\,000 while maintaining high throughput. The detector is a HAWAII-1 1024 $\times$ 1024 infrared array, cooled to 85K, with the spectrograph operating at room temperature. A shortpass filter at 85\,K, positioned in front of the detector, filters out longer infrared wavelengths. Thanks to its relatively high throughput, NIGHT on a 2-m class telescope is predicted to be as sensitive as existing high-resolution spectrographs on 4-m class telescopes.

The front end unit injects starlight and sky background into two separate optical fibers leading to the spectrograph. It also performs near-infrared guiding and includes a mechanism for injecting calibration light into either fiber.

The assembly and optical alignment of NIGHT's spectrograph and front end unit are scheduled for July-September 2024, with the first light anticipated before early 2025. Following commissioning, NIGHT is expected to begin its baseline survey, requiring 75 nights per year. 
\end{abstract}



\section{SCIENCE WITH NIGHT} \label{sec:intro}  
High-resolution (R$>$50\,000) spectroscopy has proven to be a very successful tool in exoplanet atmospheric characterization from the ground.\cite{saeger2000} The very first exoplanet atmosphere detections have been performed with space-based instrumentation (most notably HST/STIS and Spitzer ), for example for HD209458\,b in the sodium doublet\cite{charbonneau2002} and Lyman-$\alpha$\cite{vidal2003extended}, and for HD189733b by detection of dayside thermal emission\cite{drake2006}. The first ground-based confirmation of such an atmosphere came later using high-resolution observations of the sodium doublet in transits of HD209458b\cite{redfield2008}. After these first ( ground-based ) detections, the number of detected molecules in exoplanet atmospheres has quickly grown. Some examples of detected species are water (e.g.\cite{tsiaras2019}), carbon monoxide (e.g.\cite{konopacky2013}), carbon dioxide (e.g.\cite{Ahrer2023}), methane (e.g., \cite{bell2023}), hydrogen (e.g.\cite{lecavelier2010}), and helium (e.g., \cite{allart2018}), as well as much heavier species like titanium and iron (e.g.\cite{heoijmakers2018}). With the addition of ground-based, near-infrared instruments, a whole range of atmospheric markers has come within reach of the exoplanet community. 

One specific tracer, the metastable helium triplet at 1083.3 $\mathrm{nm}$ (He I), has been abundantly observed for a growing group of planets.\cite{oklopcic2018, spake2018, orell2024, guilluy2023, allart2023} Given helium's low molecular mass, it can be found at much higher altitudes in exoplanet atmospheres, being most abundant in the upper thermosphere and exosphere. Some planets, typically of low density, low core mass, and high insolation, show strongly inflated atmospheres, traced by strong absorption in the helium triplet during exoplanet transits. It is believed that XUV-heating from their host stars most dominantly causes this inflation \cite{owen2019}. However, in some scenarios, core-powered heating could also be one of the driving forces \cite{modirrousta2023}. Some of these planets show intense inflation and consequentially suffer strong atmospheric mass loss. This process of atmospheric erosion likely plays an important role in the evolution of planets. Together with planet migration, it could explain the formation of the Neptunian Desert\cite{lecavelier2007,mazeh2016} and radius valley\cite{fulton2017}, two imprints of planetary evolution in the observed demographics of planets. \cite{owenwu2017,owenlai2018,naho2022} As lined out in \cite{farretjentink2024}, current ground-based, near-infrared, high-resolution spectrographs will not be able to survey the statistics of He evaporation rates and the temporal variability of He I absorption in a large enough planet sample to address fundamental questions: the formation of the Neptunian desert and radius valley, and the general interactions between stars and close-in planets. As such, the NIGHT instrument was proposed.

\section{Description of NIGHT} \label{sec:something}
The NIGHT instrument is a fibre-fed, near-infrared, partially stabilized spectrograph, operating in a narrow wavelength regime of 1.081\,$\mu$m to 1.085\,$\mu$m. It will be the first ground-based spectrograph fully dedicated to exoplanet atmospheric characterization, aiming to survey over 100 planets in total and perform follow-up observations for dozens of them. The narrow wavelength range of NIGHT was defined at an earlier stage and chosen to accommodate the metastable He triplet absorption lines as a tracer of inflated and escaping atmospheres\cite{farretjentink2024}. NIGHT will operate at a spectral resolution of R $\simeq$ 75\,000, which was chosen to fully resolve the lineshape of the He triplet\cite{farretjentink2024}. 

The NIGHT instrument consists of the spectrograph, the fiber guiding unit, allowing starlight injection at the telescope focal plane, and a calibration and control unit. With its high throughput, NIGHT has been designed to perform its science from moderate 1.5 to 2-meter class telescopes, allowing flexibility in its observing program. Once operational, NIGHT will be the first spectrograph to perform exoplanet atmospheric characterization from a ground-based telescope of sub-2-meter aperture. The next chapters outline the details of the main components of the NIGHT instrument. 

\subsection{The spectrograph}
The spectrograph has been designed to maximize efficiency while remaining fairly compact and cost-efficient. As such, the number of optical surfaces was minimized, utilizing mostly off-the-shelf (OTS) components. For custom optics, (anti)-reflective coatings were optimized for our narrow wavelength regime. The main disperser, typically the source of the highest losses in high-resolution spectrographs, is a volume-phase-holographic grating (VPHG) in the case of NIGHT. It was custom-built by Wasatch Photonics and operates in first-order with a diffraction efficiency of $\sim$90\%. The spectrograph operates at a spectral resolution of 75\,000 and sampling of 2.9-pixel per resolution element. Due to a lack of cross-dispersion and the narrow wavelength range, we expect no variation in sampling over wavelength. NIGHT requires thermal stability, so a temperature-controlled thermal insulation box will enclose the vacuum tank.  

\subsubsection{The vacuum tank}
\begin{figure}[!h]
    \begin{center}
    \begin{tabular}{cc}
    \includegraphics[height=6cm]{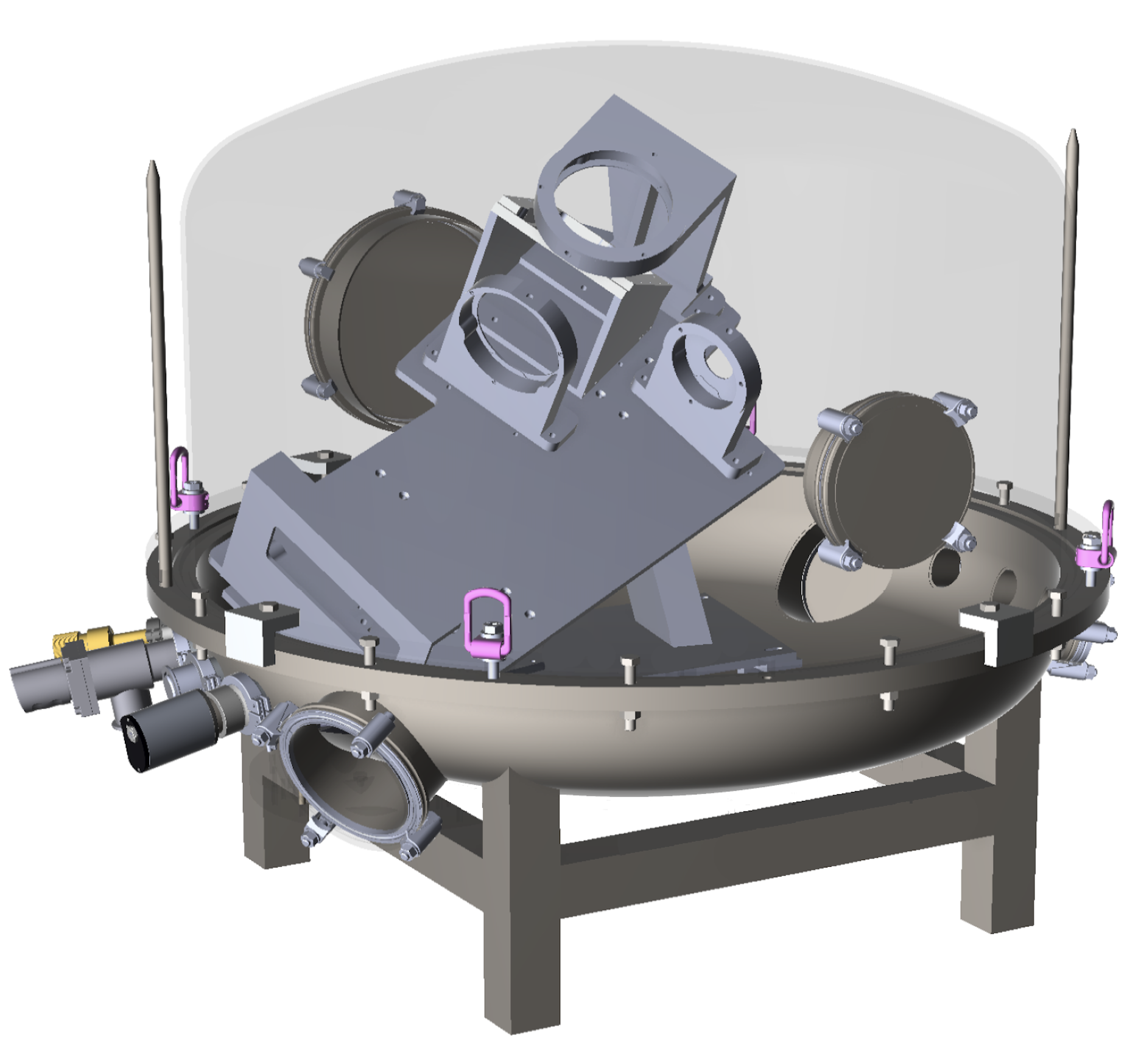}&
    \includegraphics[height=4.5cm]{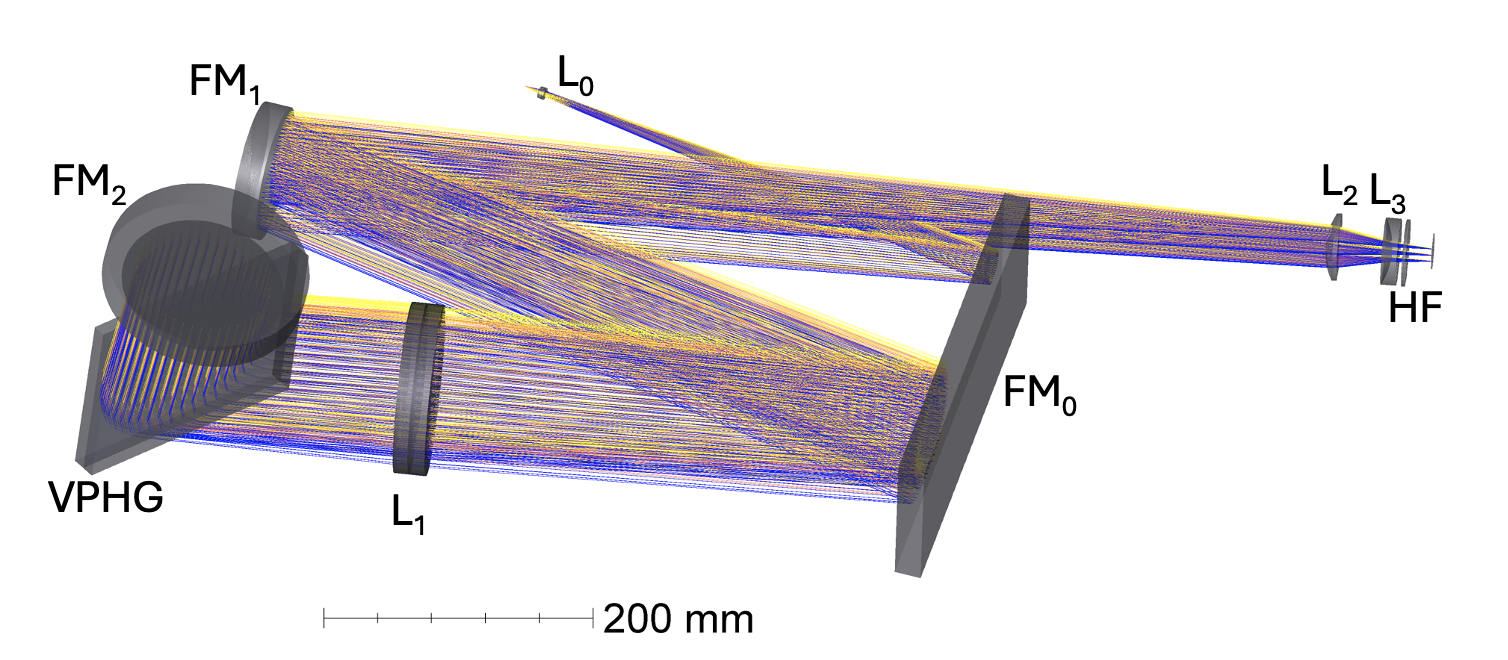}\\
    \end{tabular}
    \end{center}
    \caption[example] 
    { \label{fig:design}On the left-hand side the vacuum enclosure of NIGHT with its opto-mechanical structure inside. On the right-hand side, we find the optical ray trace of the spectrograph. On top, we find the fiber injection with a small lenslet (L0) acting as a focal extender. The beam has a first pass on the large rectangular flat (FM0) after which it folds to the circular optical flat (FM1). After a second pass on FM0, the beam hits the doublet collimator (L1). The beam passes through the VPHG twice using a second circular optical flat (FM2). Then, the beam travels back through L1, to FM0 and FM1, and towards the camera optics. L2 is a 50\,mm focal length singlet and L3 is a $-50$\,mm focal length singlet field flattened. After these lenses we find the 85\,K cold heat filter (HF).}
    \end{figure}

The vacuum tank of NIGHT is used to keep the air pressure around the optics constant. This will ensure the refractive index of the surrounding air remains constant, improving the wavelength stability of the observed spectrum. Important to note here is that we do not aim to place the optics of NIGHT under vacuum. The vacuum enclosure's purpose is solely to increase instrumental stability. A spare vacuum enclosure at the Observatory of Geneva was repurposed for cost efficiency. Figure~\ref{fig:design} shows the mechanical design of the spectrograph. Note that the optical bench is tilted inside the vessel such that one of the lower flanges on the tank can be used to mount the detector and maintain optical alignment whenever the vacuum vessel is opened up.

\subsubsection{The optical layout}

The optical layout of NIGHT was designed to allow for the use of OTS components for some parts. The camera lenses, focal extender at the fiber input, and two of the three fold mirrors are OTS. These components allowed us to reduce the total instrument cost. The VPHG, large rectangular fold mirror (FM0), and collimator (L1) are custom components. FM0 is a single component used in a triple pass setup, instead of using separate mirrors, to increase the stability and rigidity of the overall optical setup, similar to UVES and HARPS(-S, -N, 3)\cite{dekker2000,pepe2000,thompson2016}. Note that for the instruments mentioned above, the triple pass mirror was of parabolic shape, acting as the main collimator. For NIGHT the mirror is flat and collimation is achieved through the doublet L1.

For all custom components, the reflective and anti-reflective (AR) coatings were optimized for NIGHT's narrow wavelength band, resulting in losses of $<$0.1\% at each surface. The OTS lenses have a standard broadband coating supplied by the manufacturer and a theoretical reflection loss of 0.4\% for each surface. Ray trace analyses and measurements show a theoretical throughput value for the full spectrograph of 71\%, uniform over the entire bandwidth, where most losses come from the VPHG. In Figure~\ref{fig:design} we find the ray trace of the optical design.
   
\subsubsection{The double-pass VPH grating}

Given NIGHT's narrow wavelength range, cross-dispersion is unnecessary to achieve the required resolution of R $\simeq$ 75\,000 over the full bandwidth. Additionally, an echelle-grating-based solution combined with an order-sorting mechanism would result in significantly lower efficiency\cite{farretjentink2024}. Instead, a Volume Phase Holographic Grating (VPHG) was optimized for first-order diffraction efficiency. This grating, with 1407 lines per millimetre, operates at an incidence angle of $49.6^\circ$. To achieve the desired spectral resolution with a 60\,$\mu$m fibre, the required clear aperture (CA) is a substantial $180 \times 120$\,mm.

The grating was custom-designed and manufactured by Wasatch Photonics, and it is anti-reflection (AR) coated for the specified wavelength range and angle of incidence. Manufacturer tests indicate an average single-pass, pre-AR coating diffraction efficiency of 82\%, measured uniformly across five distinct zones on the grating and consistent across the wavelength range. After applying the AR coating, the diffraction efficiency is estimated to be approximately 90\%, independent of polarization and wavelength.

    \begin{figure}[!h]
    \begin{center}
    \begin{tabular}{ccc}
    \includegraphics[width=0.3\textwidth]{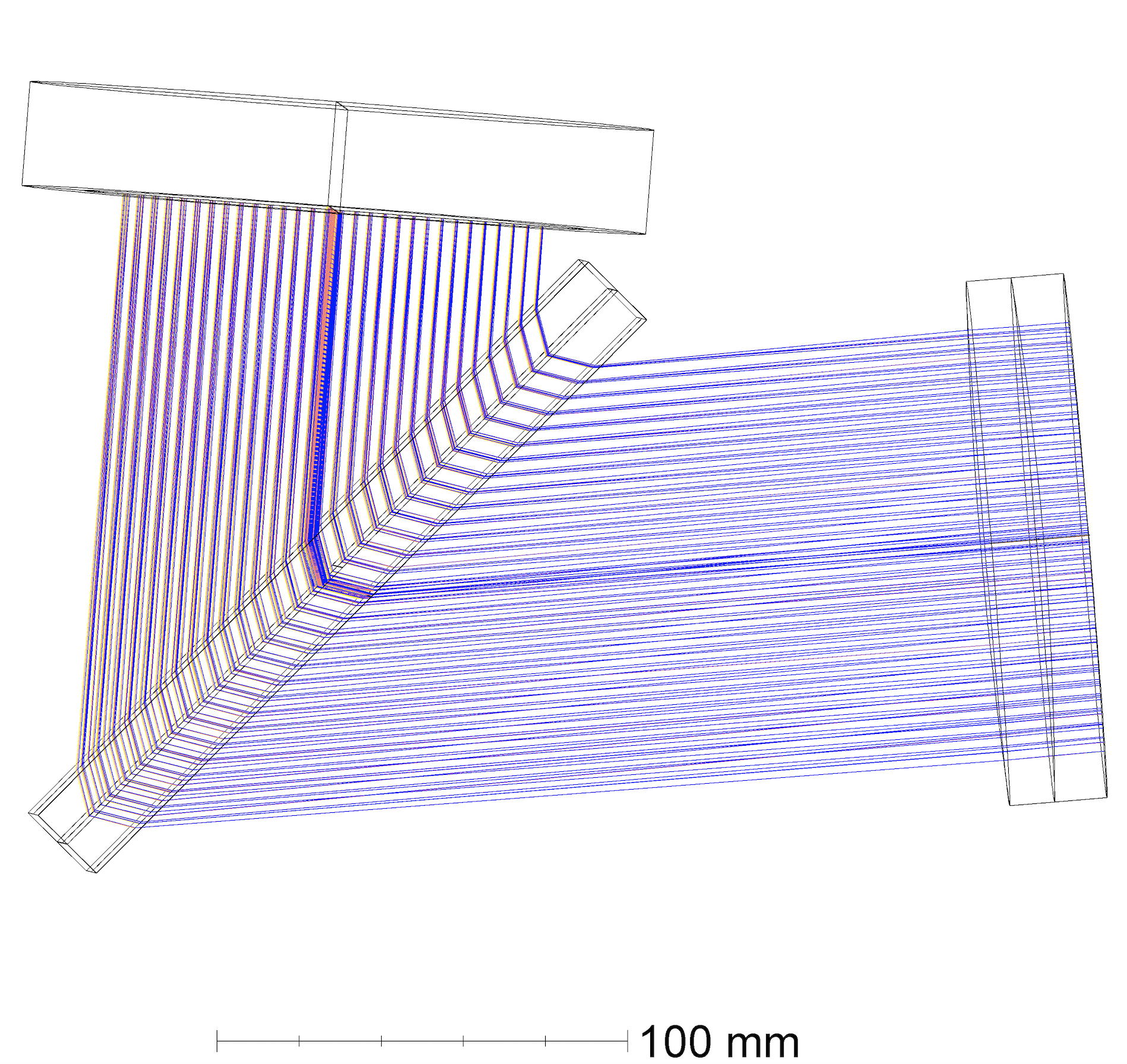}&
    \includegraphics[width=0.25\textwidth]{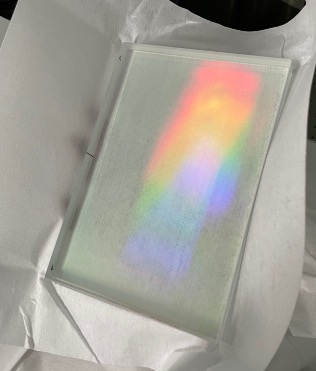}&
    \includegraphics[width=0.4\textwidth]
    {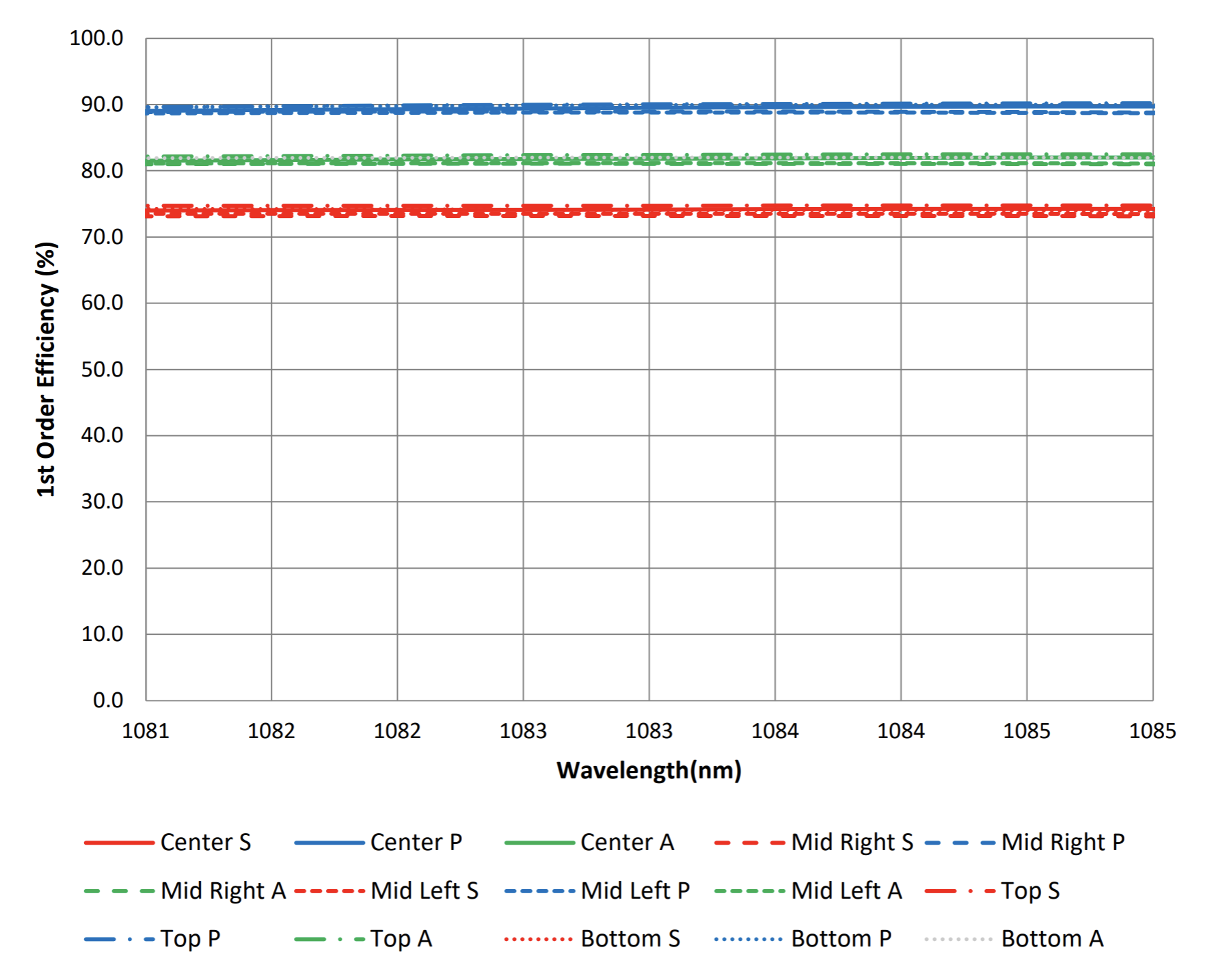}\\
    \end{tabular}
    \end{center}
    \caption[example] 
    { \label{fig:VPH} On the left-hand side we find the optical layout of the double-pass VPH grating in NIGHT. In the center is a picture of the VPH as manufactured by Wasatch Photonics. On the right-hand side are the diffraction efficiencies measured before AR coating. The P-polarized efficiency is around 89\%, whereas the S-polarized efficiency is around 73\%. After AR coating, the diffraction efficiency should be around 90\% for both polarizations.}
    \end{figure}

\subsection{The detector}
Given NIGHT operates at nIR wavelengths, it utilizes a HAWAII series HgCdTe detector array. The \hbox{HAWAII-1} 1024 $\times$ 1024 array that will be used for NIGHT was repurposed from the TRIDENT instrument, a nIR imager that was placed at the Observatoire du Mont M\'egantic (OMM) and the Canada-France-Hawaii Telescope (CFHT)\cite{marois2005}. The detector is housed in an Infrared Laboratories liquid nitrogen (LN2) cooled bath cryostat. 

\subsubsection{Detector performance}
\begin{figure}[h]
    \centering
    \includegraphics[width=0.95\textwidth]{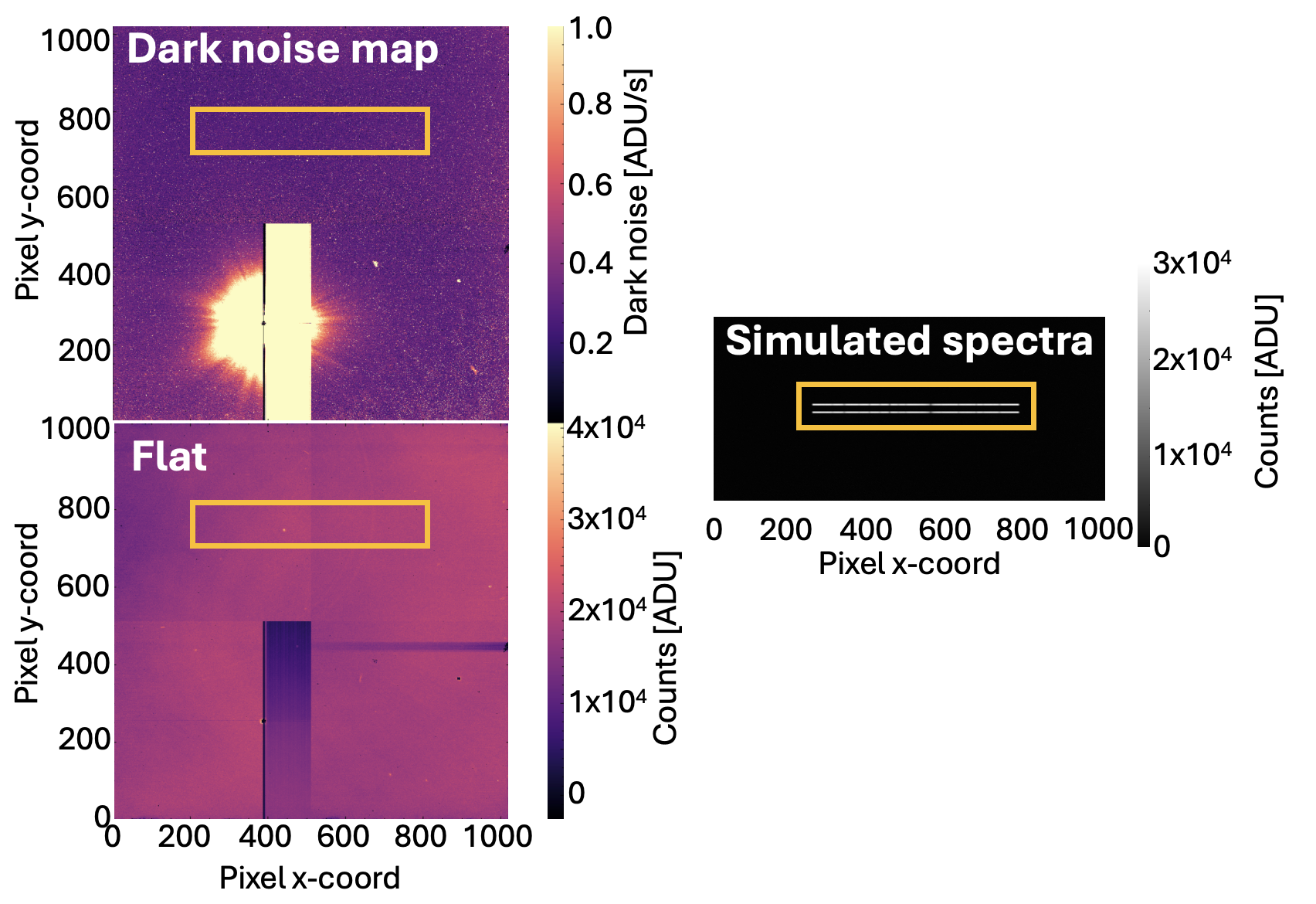}
    \caption{The detector performance of NIGHT. The dark noise map was computed through dark measurements over increasing exposure time and performing a linear regression per pixel. The orange box shows the proposed detector region to place the two spectra from the two channels of NIGHT. For illustrative purposes, two stellar spectra are currently simulated. In reality, one channel would transmit the star, and the other channel the sky during science operations. }
    \label{fig:det_perf}
\end{figure}

Detector testing shows good performance for two adjacent quadrants of the array. NIGHT requires a footprint of $\sim$700 $\times$ 30 pixels, so bad performance for half of the array is of no concern. In Figure~\ref{fig:det_perf} we find the detector performance in terms of dark noise, a flat field, and a simulated measurement of two stellar spectra based on these measurements.

\subsection{FIGU2: The front end unit} 

\begin{figure}[!h]
    \begin{center}
    \begin{tabular}{ccc}
    \includegraphics[height=5cm]{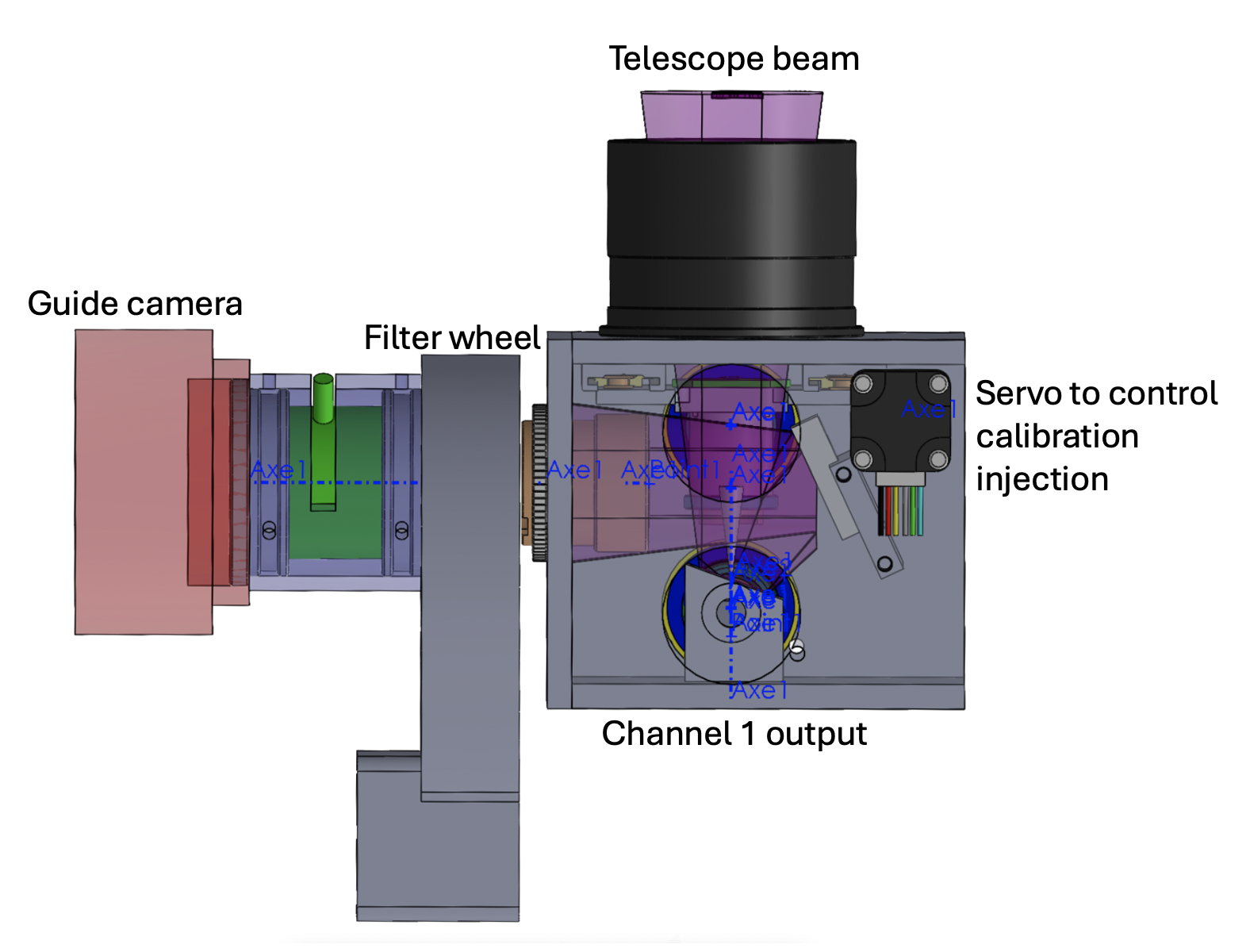}&
    \includegraphics[height=5cm]{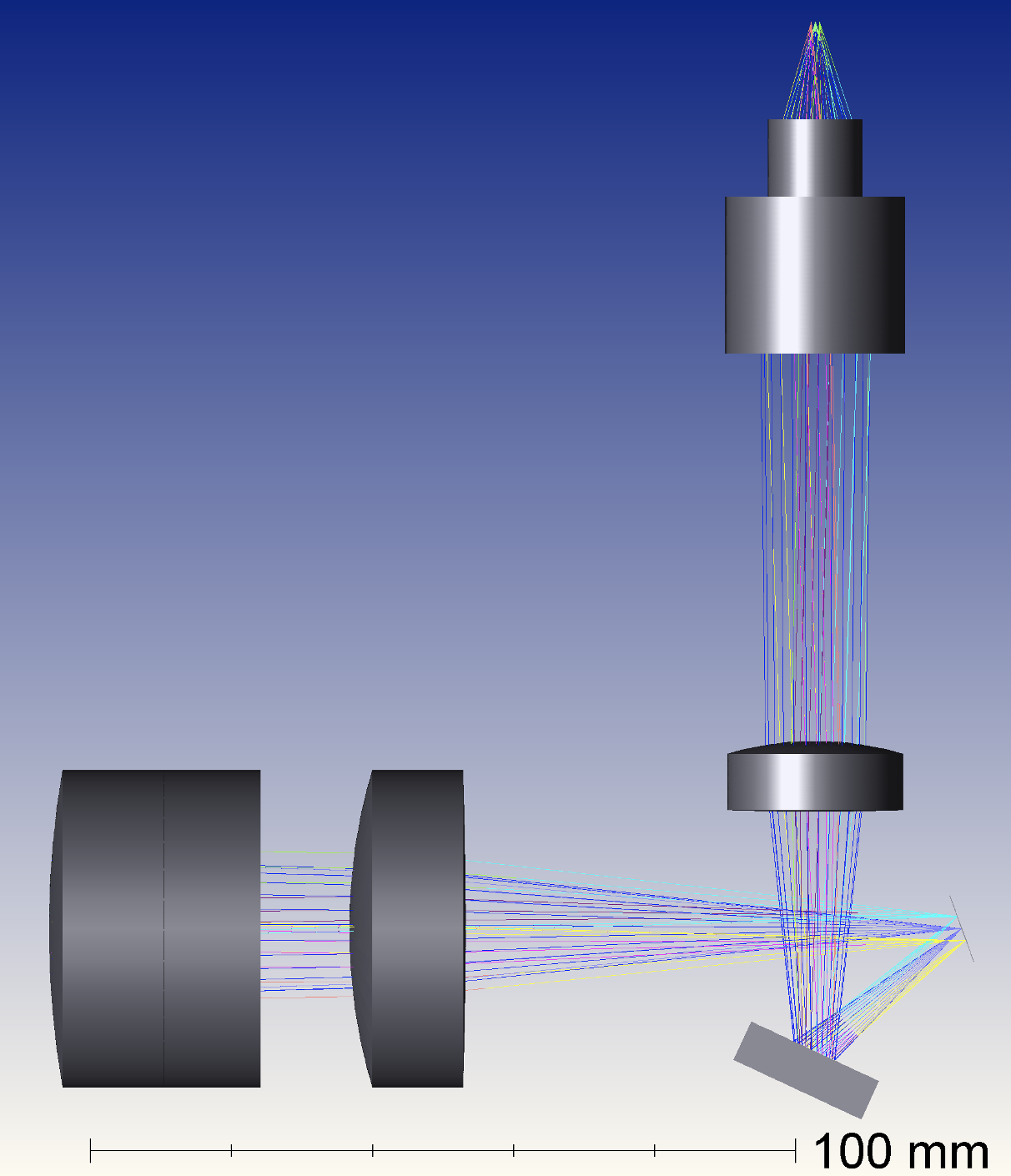}&
    \includegraphics[height=5cm]{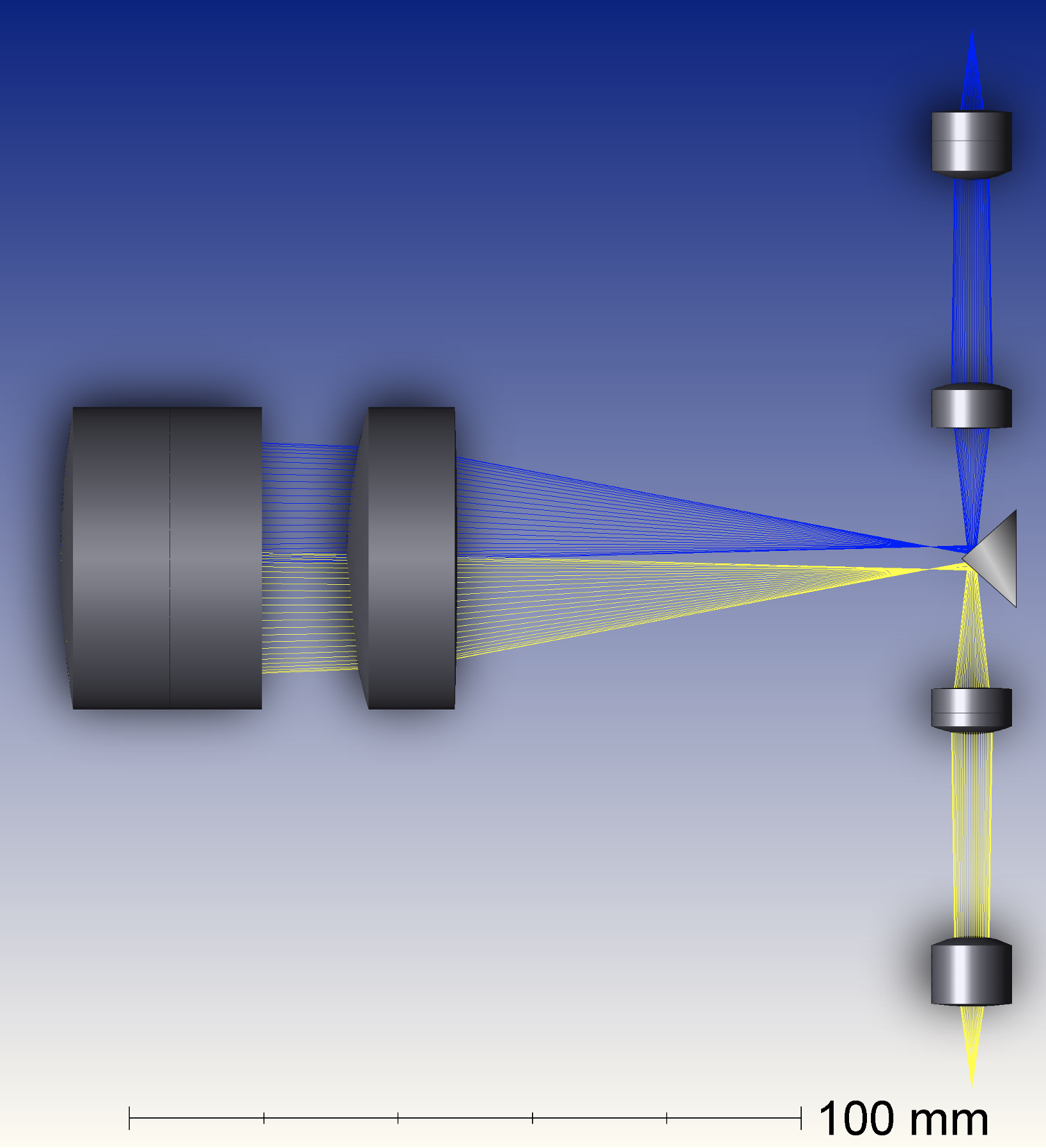}\\
    \end{tabular}
    \end{center}
    \caption[example] 
    { \label{fig:frontend_design}The design and optical ray tracing of the front end unit of NIGHT are illustrated in the accompanying figures. The mechanical design of the front end unit is depicted on the left-hand side. Telescope light enters the unit from the top, while calibration light can be introduced through an upper FC/PC fiber connector on the backside of the module (not visible in this diagram). A stepper-motor-powered slider mechanism positions the injected light above either of the two pierced holes, allowing calibration light to be fed into the fibers individually, though not simultaneously. The angled pierced mirror and a secondary fold mirror direct light toward NIGHT's guide camera.

    The center image presents the ray trace of the front end towards the guide camera. After passing through the custom focal reducer (left-hand-side), light reflects off the pierced mirror, then off a secondary flat mirror, and subsequently passes through changeable filters in a filter wheel before reaching the guide camera optics. The holes in the first mirror channel light towards the injection optics.

    The figure on the right-hand side illustrates the ray trace of the injection optics. The configuration of two lenses facilitates the conversion of an F/5 beam to an F/4 beam, required for injection into the spectrograph.}
    \end{figure}

The front end unit of NIGHT (FIGU2: FIber Guide Unit 2) has two main purposes: it injects starlight, calibration light, or the sky background into two optical fibers, and it takes guiding images at a short interval. The front end unit of NIGHT has been developed in partnership with Shelyak Instruments and is designed to be flexible and adaptable. For example, it can adapt to different telescope sizes, F-ratios, operational wavelengths, and fiber sizes. Figure~\ref{fig:frontend_design} shows schematics of the unit.

The front end unit has two output channels. Both channels are symmetric and inject light into two 60\,$\mu$m fibers at F/4 (0.125NA). For a 1.5-meter telescope, this translates to $\sim$$2^{\prime\prime}$ on the sky, and for a 2-meter telescope to $\sim$$1.55^{\prime\prime}$. The front end unit can operate in various modes, outlined in Table~\ref{Table:FIGU_modes}. For NIGHT, the calibration procedure is foreseen to take calibration exposures of a Uranium-Neon lamp for wavelength calibration, in both channels separately. More calibration exposures are taken for flat fielding, by illumination with a Tungsten lamp, also for both channels individually. We have designed the front end unit so that calibration light cannot be injected into two fibers simultaneously. Also, simultaneous wavelength calibration during science exposures is not possible. Both modes are typically required for high-precision radial velocity measurements and, as such, are unnecessary for the science that NIGHT will conduct. These simplifications in the operational requirements for NIGHT's front end unit allowed for design simplifications and overall size and complexity reduction. Science exposures are taken when starlight is injected into either channel 1 or 2. The other channel will then sample the sky background, used in post-processing for telluric correction. The separation of the two channels is 3\,mm at an F/5 focus in the front end unit. This translates to an angular separation of 1.375$^\prime$ on the sky for a 1.5-meter telescope.
Additionally, the front end unit has a dark mode, allowing the user to take dark exposures by fully blocking the incoming telescope beam, essentially acting as a shutter.

\begin{table}[ht]
\caption{Operational modes of the front end unit of NIGHT.} 
\label{Table:FIGU_modes}
\begin{center}       
\begin{tabular}{l|l|l} 
\hline
\rule[-1ex]{0pt}{3.5ex}  \textbf{Mode} & \textbf{Channel 1} & \textbf{Channel 2}  \\
\hline
\rule[-1ex]{0pt}{3.5ex}  Calibration 1 & Uranium-Neon or Tungsten-Halogen & empty  \\
\hline
\rule[-1ex]{0pt}{3.5ex}  Calibration 2 & empty & Uranium-Neon or Tungsten-Halogen \\
\hline
\rule[-1ex]{0pt}{3.5ex}  Science 1 & Star & Sky  \\
\hline
\rule[-1ex]{0pt}{3.5ex}  Science 2 & Sky & Star \\
\hline
\rule[-1ex]{0pt}{3.5ex}  Dark & empty & empty \\
\end{tabular}
\end{center}
\end{table} 

\subsubsection{Focal reducer and guiding}
NIGHT has been foreseen to function as a visitor instrument, possibly changing observatories over its lifetime. As such, the front end unit has been designed to be flexible and adaptable to different telescope focal ratios. The fiber size and numerical aperture (NA) of injection are fixed. Also, the optics in the front end accept an F/5 focus from the telescope. Any telescope focus not at F/5 thus requires a focal reducer (in case of a slower beam), or focal extender (in case of a faster beam). The incoming telescope beam in Figure~\ref{fig:frontend_design} is a F/30 Coude focus from a 1.5-meter telescope. The focal reducer designed for this telescope is a 3-element system consisting of a doublet and singlet. In the current design, it has been foreseen that the whole system is placed about 50cm from the original Coude focus. The guide camera optics consist of a 70\,mm doublet for collimation and a 7-element camera with a focal ratio of F/1.8. 

To increase the flexibility of the system, the front end does not employ an atmospheric dispersion corrector (ADC). Instead, guiding in the near-infrared is achieved by implementing a longpass filter with a cut-on wavelength of 850 nm in front of the guide camera. The guide camera used is a ZWO ASI462MM, which is sensitive in the near-infrared with a cutoff in sensitivity at 1100\,nm. At an airmass of 2, the atmospheric dispersion from 850 nm to 1100 nm is approximately $0.4^{\prime\prime}$, equivalent to 1/5$^{\rm th}$ of our fiber diameter or 2 pixels on the guide camera for a 1.5-meter telescope. This dispersion results in a slight blurring of the guide camera image, which captures the fiber from 850 nm to 1100 nm.

Due to the non-uniform sensitivity of the guide camera across the guiding wavelength band, the blurring effect is reduced. However, since the sensitivity is higher at shorter wavelengths, which are farther from NIGHT's operational wavelength band, some reduction in fiber injection efficiency may occur during high airmass observations. Further testing of the exact throughput values over the wavelength range for the guiding optics is necessary to determine if implementing an airmass-dependent offset in the pointing corrections provided by the guide system can improve fiber injection efficiency.

\begin{figure}[h]
    \centering
    \includegraphics[width=0.9\textwidth]{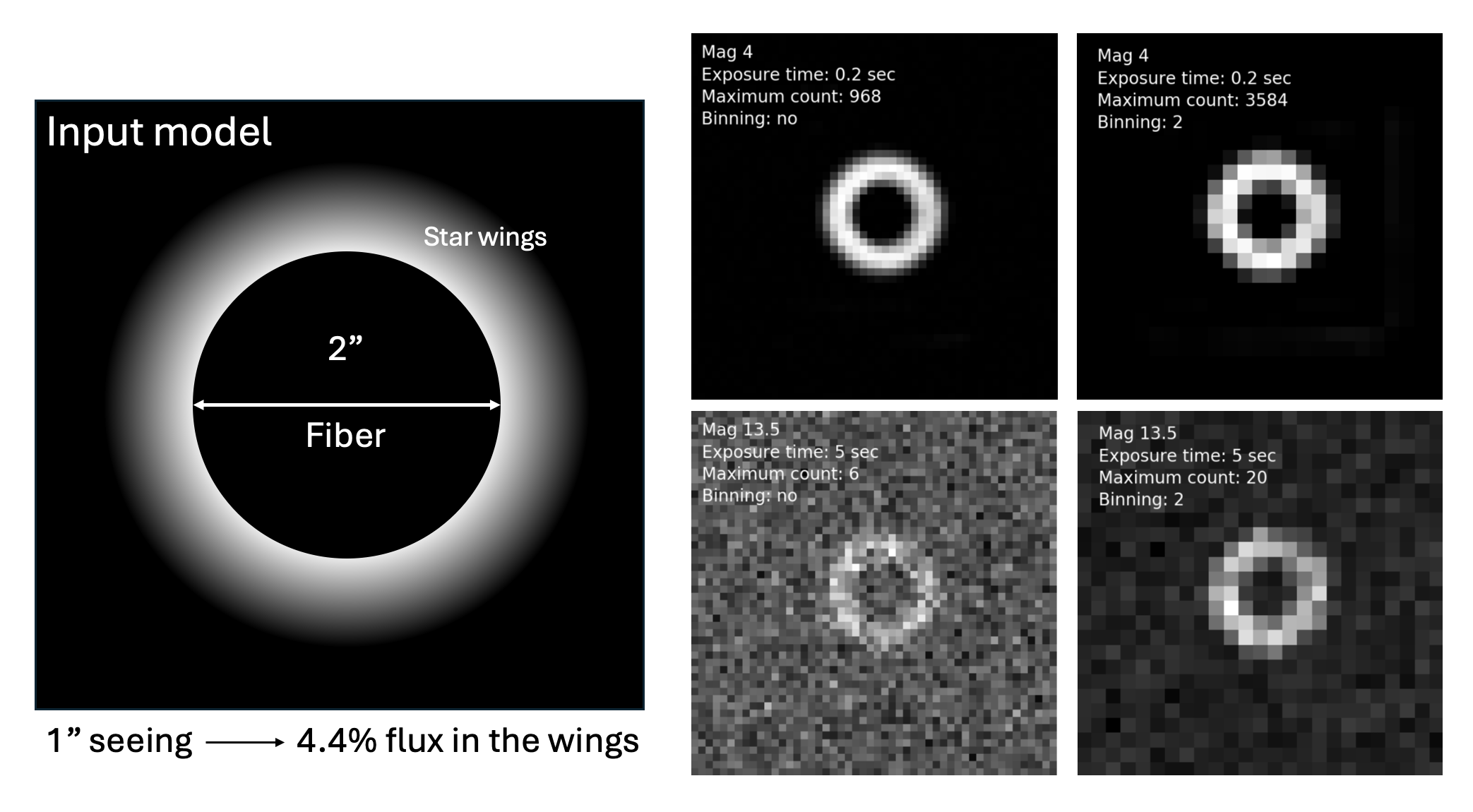}
    \caption{The pierced hole for fiber injection as seen by the guide camera of NIGHT's front end unit. NIGHT does guiding by imaging the wings of the star. The 60\,$\mu$m fiber at F/4 (0.125NA) translates to 2" on the sky for a 1.5-meter telescope. Assuming a median $1^{\prime\prime}$ seeing, this translates to 4.4\% flux in the star's wings. Assuming 5\% throughput from the top of the atmosphere to the detector, including Q.E., sky emission of 5 $\mathrm{phot/nm/m^2/arcsec^2/s}$, read-out noise of 1e$^-$rms, and negligible dark current, the resulting detector images can be seen above. On the right-hand side, we applied the detector's bin2 mode, merging 4 pixels into 1, to improve SNR. We find that even at magnitude 13.5, there is still an acceptable fiber image for guiding in the bin2 mode. Initial targets for NIGHT are foreseen to be of $J$-band magnitude $<$12\cite{farretjentink2024}, so the current design seems feasible with room for margin. With proper background subtraction, we anticipate being able to reach even fainter magnitudes.}
    \label{fig:guiding}
\end{figure}

To determine the limiting magnitude of the guide camera system, we performed a Zemax optical raytrace simulation of the guide system, generating detector images for various magnitudes and integration times. The results of these simulations are presented in Figure~\ref{fig:guiding}.

\subsection{The calibration and control unit}
\begin{figure}[h]
    \centering
    \includegraphics[width=0.3\textwidth]{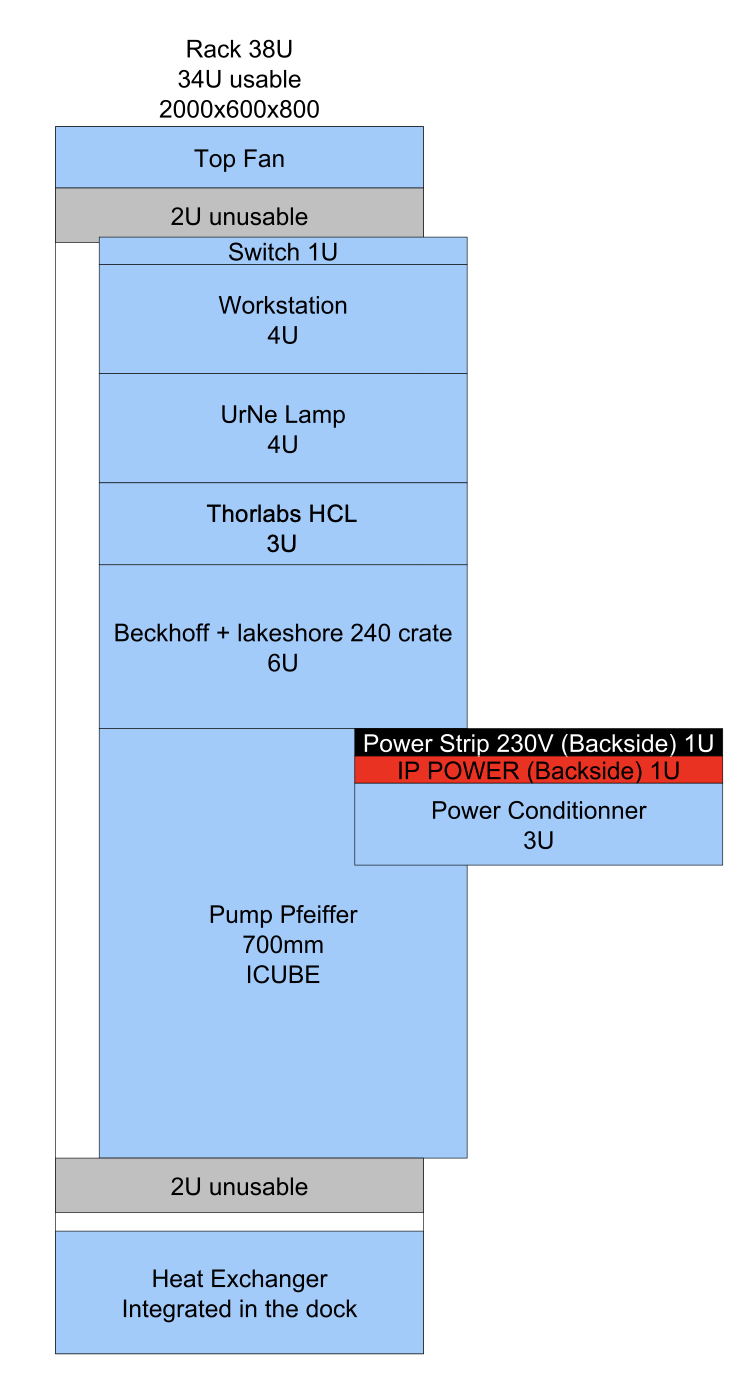}
    \caption{The calibration and control unit of NIGHT. Composed of a single rack it houses from bottom to top: the cryostat vacuum pump, Beckhoff control, Lakeshore 240 detector temperature control, a Thorlabs Tungsten lamp, Ur-Ne lamp, and the workstation.}
    \label{fig:calconrack}
\end{figure}

The calibration and control unit comprises a single rack housing the calibration light sources, their control systems, and the instrument control hardware. A simplified diagram of the rack organization can be found in Figure~\ref{fig:calconrack}. The calibration sources for NIGHT are a Uranium-Neon lamp for wavelength calibration and a Tungsten-Halogen lamp for flat field calibration. Both sources and their optomechanics for fiber injection are fully OTS. 

\subsection{The fiber-link}
The fiber-link system of NIGHT was designed to prioritize simplicity. To minimize the complexity of control systems, the individual fibers from the calibration sources were spliced into a single calibration fiber. Specifically, the fiber combiner merges three 100\,$\mu$m fibers into a single 200\,$\mu$m fiber. Although two calibration sources are currently utilized, the design allows for an additional source. Calibration source selection is achieved by activating or deactivating the individual light sources. Each calibration source is equipped with a neutral density (ND) filter slot to adjust flux levels appropriately, ensuring the desired signal-to-noise ratio (SNR) in calibration exposures.

The 200\,$\mu$m calibration fiber directs light to the front end unit, where it can be injected into either Channel 1 or Channel 2. The front end unit injects light into both channels at F/4. Channels 1 and 2 consist of 60\,$\mu$m circular fibers that lead to the spectrograph. Two meters before the double scrambler unit, the circular fibers transition to 57.5\,$\mu$m octagonal fibers (with an effective diameter of 60\,$\mu$m). The double scrambler units, in combination with the octagonal fibers, scramble the signal in both near-field and far-field\cite{Bouchy1999}. Each channel has its own double scrambler. In addition to scrambling, each double scrambler houses two optical filters: one shortpass and one longpass filter with cut-on and cut-off wavelengths of 1050 nm and 1100 nm, respectively. These filters eliminate most light outside NIGHT's operational band, reducing contamination from scattered light within the spectrograph. After passing through the double scramblers, the two channels are combined into a single ferrule, with the octagonal fiber cores separated by a 200\,$\mu$m gap. A small double lens after the fibers reduces the focal ratio to the appropriate level for the spectrograph.

To minimize coupling losses, all air-gap interfaced fiber surfaces from the front end to the spectrograph injection point are coated with a nanostructured moth-eye antireflective coating, achieving reflection losses of approximately 0.01\% at each surface. All fiber-to-fiber interfaces are coupled using index-matching gel.

\begin{figure}[h]
    \centering
    \includegraphics[width=0.75\textwidth]{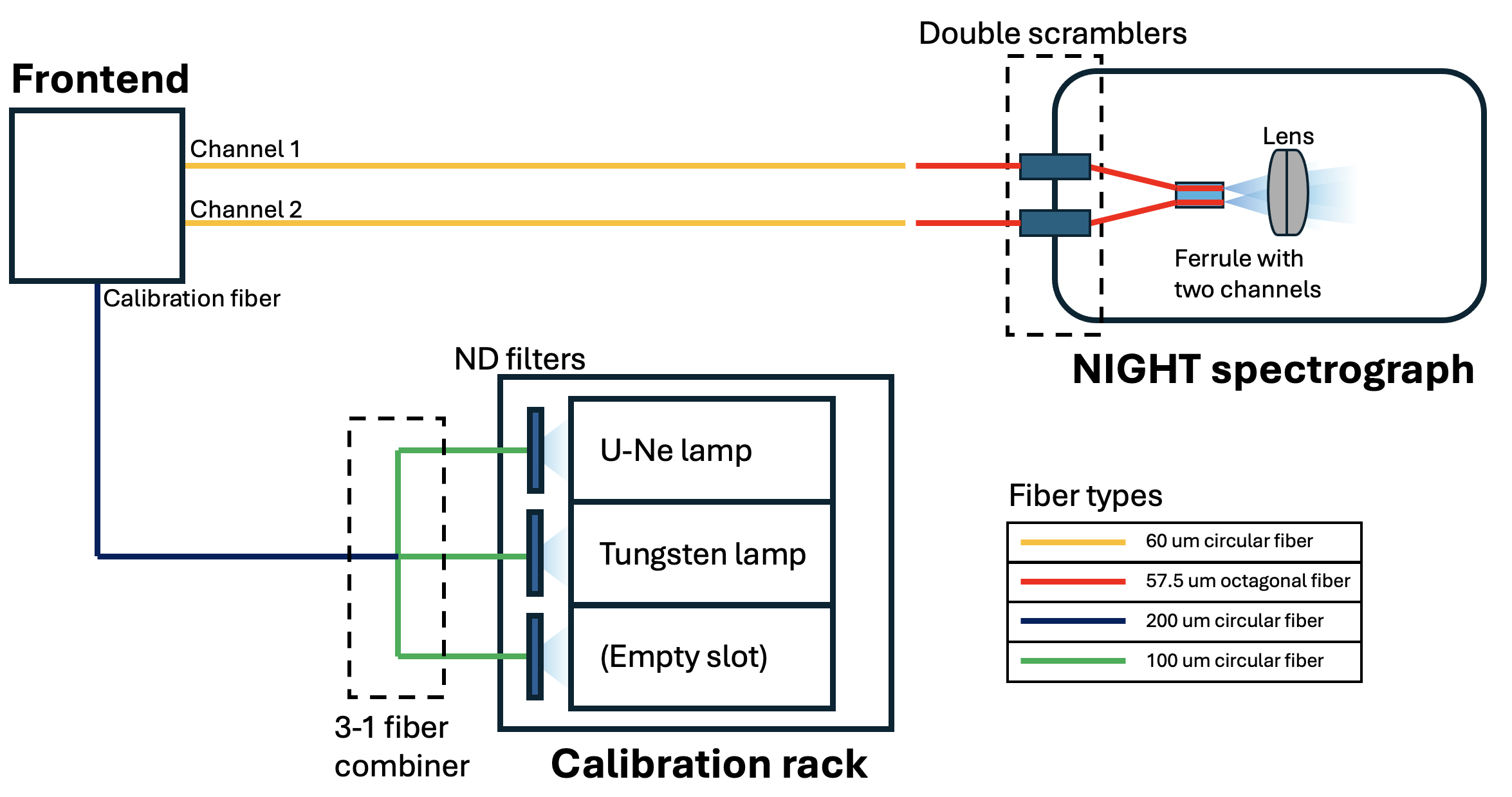}
    \caption{The full fiber train of NIGHT from calibration and control unit to spectrograph. The four different fiber types are color-coded. }
    \label{fig:fiber_train}
\end{figure}

\subsection{Expected performance of NIGHT}

\subsubsection{System throughput}
\begin{figure}[h]
    \centering
    \includegraphics[width=0.5\textwidth]{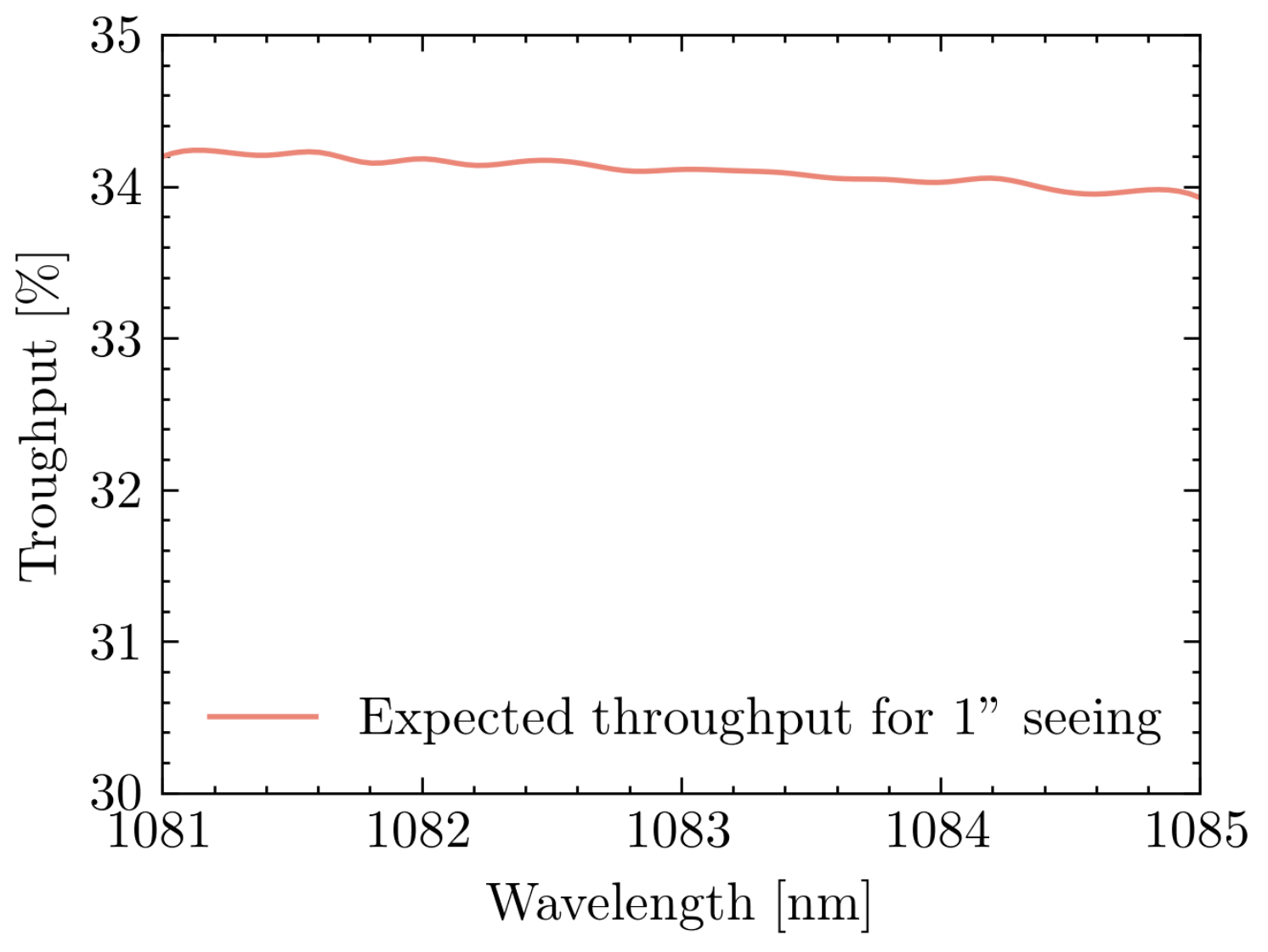}
    \caption{The total throughput of NIGHT over its operational bandwidth from 1081 to 1085nm. The throughput is close to uniform over the entire wavelength band resulting from the VPH grating's uniform diffraction efficiency. High-frequency oscillations in the transmission curve are induced by the dielectric coatings of the filters in the double scrambler and the heat filter inside NIGHT's cryostat. }
    \label{fig:throughput}
\end{figure}

The total system throughput is plotted in Figure~\ref{fig:throughput}. The average throughput is estimated to be at the level of 34\%. This efficiency is significantly higher than for previously built high-resolution near-infrared spectrographs like SPIRou and NIRPS, which have throughputs on the order of 4-13\%, depending on wavelength and observing mode \cite{bouchy2017, boucher2021}. The reported throughput values are typically the maximum values and can vary significantly between wavelength bands and over a spectral order. As a result, NIGHT is expected to achieve similar scientific results in the helium triplet regime as aforementioned instruments while operating on a significantly smaller telescope. The transmission curve of NIGHT accounts for the throughput of the:
\begin{enumerate}
    \item Atmosphere (95\%)
    \item Telescope (70\%)
    \item Seeing (1", 95.6\%)
    \item Injection (92\%)
    \item Double scrambler and fibers ($\sim$82\%)
    \item Spectrograph ($\sim$71\%)
\end{enumerate}

\section{Outlook}
As of this writing, the NIGHT spectrograph and its front end have successfully passed the Final Design Review (FDR). The manufacturing of NIGHT's mechanical components for both the spectrograph and front end is scheduled for July and August 2024. Concurrently, the assembly and alignment of these components will be conducted at the Observatory of Geneva, with completion anticipated by late September. During the same period (July-September), we also plan to assemble the calibration and control rack.

Post-September, NIGHT will be transferred to a 1.5-meter telescope for commissioning, which is expected to occur during the final months of 2024. In 2025, we aim to commence our baseline observing program, targeting 75 nights per year, with the potential for additional observing nights. The specific telescope to be utilized for this program has not been definitively selected, though several options are under consideration. In the longer term, there is interest in relocating NIGHT to a 2-meter telescope to fully leverage its scientific capabilities. 

\acknowledgments
The NIGHT instrument is the result of a collaboration between the Department of Astronomy at the University of Geneva (also named the Observatory of Geneva), the Observatoire du Mont-Mégantic, the Trottier Institute of the Research of Exoplanets at the Université de Montréal, and the NCCR PlanetS, the interdisciplinary Swiss planetary research program comprising the Universities of Geneva, Bern, Zurich, and ETH Zurich. NIGHT is funded by the Swiss National Science Foundation (SNSF) grants 184618, 51NF40182901, and 51NF40205606. NIGHT is also supported by a grant from the \textit{Fonds de recherche du Québec – Nature et technologies} (FRQNT) and the Institut Trottier de Recherche sur les Exoplanetes (IREX). R.A. acknowledges the SNSF support under the Post-Doc Mobility grant P500PT-222212 and the support from IREX.

\bibliography{report}   
\bibliographystyle{spiebib}   

\end{document}